\documentclass{article}
\usepackage{cite}
\usepackage{geometry}
\usepackage{color}
\usepackage{amsmath}
\usepackage{wasysym}
\usepackage{amssymb}
\usepackage{ulem}
\usepackage{listings}
\definecolor{mygreen}{rgb}{0,0.6,0}
\usepackage{tabularx}
\begin{document}
\title{Theory Manual for the Tuned Mass Damper Module in FAST v8}  
\author{William La Cava \& Matthew A. Lackner \\
Department of Mechanical and Industrial Engineering\\
University of Massachusetts Amherst\\
Amherst, MA 01003 \\
\texttt{wlacava@umass.edu}, \texttt{lackner@ecs.umass.edu}} 
\maketitle

This manual describes new functionality in FAST 8 that simulates the addition of tuned mass dampers (TMDs) in the nacelle for structural control. For application studies of these systems, refer to \cite{lackner_passive_2011, lackner_structural_2011,namik_active_2013, stewart_effect_2011,stewart_impact_2014,stewart_optimization_2013}. The TMDs are two independent, 1 DOF, linear mass spring damping elements that act in the fore-aft ($x$) and side-side ($y$) directions. We first present the theoretical background and then describe the code changes. 
\section{Theoretical Background}
\subsection{Definitions}

\begin{itemize}
\item[] $O$: origin point of global inertial reference frame 
\item[] $P$: origin point of non-inertial reference frame fixed to nacelle where TMDs are at rest
\item[] $TMD$: origin point of a TMD
\item[] $G$: axis orientation of global reference frame
\item[] $N$: axis orientation of nacelle reference frame with unit vectors $\hat{\imath}, \hat{\jmath}, \hat{k}$
\item[] $\vec{r}_{_{_{TMD/O_G}}} = \left[ \begin{array}{c} x \\ y\\ z \end{array} \right]_{_{TMD/O_G}}$: position of a TMD with respect to (w.r.t.) $O$ 
with orientation $G$  
\item[] $\vec{r}_{_{_{TMD/P_N}}} = \left[ \begin{array}{c} x \\ y\\ z \end{array} \right]_{_{TMD/P_N}}$: position of a TMD w.r.t. $P_N$ 
\item[] $\vec{r}_{_{_{TMD_X}}}$: position vector for $TMD_X$ \\
\item[]$\vec{r}_{_{_{TMD_Y}}}$: position vector for $TMD_Y$ 
\item[]$\vec{r}_{_{P/O_G}} =\left[ \begin{array}{c} x \\ y\\ z \end{array} \right]_{_{P/O_G}} $: position vector of nacelle w.r.t. $O_G$   
\item[]$R_{_{N/G}}$: 3 x 3 rotation matrix transforming orientation $G$ to $N$ 
\item[]$R_{_{G/N}} = R_{_{N/G}}^T $: transformation from $N$ to $G$
\item[] $\vec{\omega}_{_{N/O_N}} = \dot{\left[ \begin{array}{c} \theta \\ \phi \\ \psi \end{array} \right]}_{_{N/O_N}}$: angular velocity of nacelle in orientation $N$; defined likewise for $G$  \\
\item[] $\dot{\vec{\omega}}_{_{N/O_N}} = \vec{\alpha}_{_{N/O_N}}$: angular acceleration of nacelle   \\
\item[]$\vec{a}_{G/O_G} = \left[ \begin{array}{c}0 \\ 0\\ -g \end{array} \right]_{/O_G}$: gravitational acceleration in global coordinates
\item[]$\vec{a}_{G/O_N} = R_{_{N/G}} \vec{a}_{G/O_G} = \left[ \begin{array}{c}a_{_{G_X}} \\ a_{_{G_Y}}\\ a_{_{G_Z}} \end{array} \right]_{/O_N}$: gravity w.r.t. $O_N$ 
\end{itemize}
\subsection{Equations of motion}
The position vectors of the TMDs in the two reference frames $O$ and $P$ are related by

 \begin{displaymath}
 \vec{r}_{_{TMD/O_G}} =  \vec{r}_{_{P/O_G}} +  \vec{r}_{_{TMD/P_G}}
 \end{displaymath}
 Expressed in orientation $N$,
  \begin{displaymath}
 \vec{r}_{_{TMD/O_N}} =  \vec{r}_{_{P/O_N}} +  \vec{r}_{_{TMD/P_N}} 
 \end{displaymath}
\begin{displaymath}
  \Rightarrow \vec{r}_{_{TMD/P_N}} =  \vec{r}_{_{TMD/O_N}} -  \vec{r}_{_{P/O_N}}
  \end{displaymath}
  Differentiating,\footnote{Note that $( R a ) \times ( Rb ) = R( a \times b )$.}
 \begin{displaymath}
 \dot{\vec{r}}_{_{TMD/P_N}}= \dot{\vec{r}}_{_{TMD/O_N}} - \dot{\vec{r}}_{_{P/O_N}} - \vec{\omega}_{_{N/O_N}} \times \vec{r}_{_{TMD/P_N}}
 \end{displaymath} 

 differentiating again gives the acceleration of the TMD w.r.t. $P$ (the nacelle position), oriented with $N$:
 
 \begin{equation} \label{accel}
 \begin{array}{cc} \ddot{\vec{r}}_{_{TMD/P_N}} = &  \ddot{\vec{r}}_{_{TMD/O_N}}  - \ddot{\vec{r}}_{_{P/O_N}} - \vec{\omega}_{_{N/O_N}} \times (\vec{\omega}_{_{N/O_N}} \times \vec{r}_{_{TMD/P_N}}) \\[1.1em] &- \vec{\alpha}_{_{N/O_N}} \times \vec{r}_{_{TMD/P_N}} - 2 \vec{\omega}_{_{N/O_N}} \times \dot{\vec{r}}_{_{TMD/P_N}} \end{array}
 \end{equation}
 
 The right-hand side contains the following terms:
 \begin{itemize}
 \item[] $\ddot{\vec{r}}_{_{TMD/O_N}}$: acceleration of the TMD in the \textit{inertial} frame $O_N$
 \item[] $\ddot{\vec{r}}_{_{P/O_N}} = R_{_{N/G}} \ddot{\vec{r}}_{_{P/O_G}}$: acceleration of the Nacelle origin $P$ w.r.t. $O_N$ 
 \item[] $ \vec{\omega}_{_{N/O_N}} = R_{_{N/G}} \vec{\omega}_{_{N/O_G}}$ : angular velocity of nacelle w.r.t. $O_N$  
 \item[] $ \vec{\omega}_{_{N/O_N}} \times (\vec{\omega}_{_{N/O_N}} \times \vec{r}_{_{TMD/P_N}})$ : Centrifugal force 
 \item[] $\vec{\alpha}_{_{N/O_N}} \times \vec{r}_{_{TMD/P_N}}$: Euler force
 \item[] $2\vec{\omega}_{_{N/O_N}} \times \dot{\vec{r}}_{_{TMD/P_N}}$: Coriolis force
 \end{itemize}

 
 The acceleration in the inertial frame $\ddot{\vec{r}}_{_{TMD/O_N}}$ can be replaced with a force balance
 \begin{align*}
  \ddot{\vec{r}}_{_{TMD/O_N}} = \left[ \begin{array}{c} \ddot{x} \\ \ddot{y} \\ \ddot{z} \end{array} \right]_{_{TMD/O_N}} = \frac{1}{m} \left[ \begin{array}{c} \sum{F_X} \\ \sum{F_Y} \\ \sum{F_Z} \end{array} \right]_{_{TMD/O_N}} = \frac{1}{m} \vec{F}_{_{TMD/O_N}}
 \end{align*}
 
 
 Substituting the force balance into Equation \ref{accel} gives the general equation of motion for a TMD:
  \begin{equation}\label{EOM}
  \begin{array}{cc} \ddot{\vec{r}}_{_{TMD/P_N}} = & \frac{1}{m} \vec{F}_{_{TMD/O_N}} - \ddot{\vec{r}}_{_{P/O_N}} - \vec{\omega}_{_{N/O_N}} \times (\vec{\omega}_{_{N/O_N}} \times \vec{r}_{_{TMD/P_N}}) \\[1.1em] & - \vec{\alpha}_{_{N/O_N}} \times \vec{r}_{_{TMD/P_N}}  - 2 \vec{\omega}_{_{N/O_N}} \times \dot{\vec{r}}_{_{TMD/P_N}} \end{array}
 \end{equation}
 We will now solve the equations of motion for $TMD_X$ and $TMD_Y$. 
 

\paragraph{TMD\_X :}

The external forces $\vec{F}_{_{TMD_X/O_N}}$ are given by 
 \begin{displaymath} \label{TMX_forces}
   \vec{F}_{_{TMD_X/O_N}} = \left[\begin{array}{c} -c_x \dot{x}_{_{TMD_X/P_N}} - k_x x_{_{TMD_X/P_N}} + m_x a_{_{G_X/O_N}} + F_{ext_x} + F_{StopFrc_{X}} \\ F_{Y_{_{TMD_X/O_N}}} + m_x a_{_{G_Y/O_N}}  \\ F_{Z_{_{TMD_X/O_N}}} + m_x a_{_{G_Z/O_N}} \end{array} \right]
 \end{displaymath}

$TMD_X$ is fixed to frame $N$ in the $y$ and $z$ directions so that
\begin{displaymath}
{r}_{_{TMD_X/P_N}} = \left[ \begin{array}{c} x_{_{TMD_X/P_N}} \\ 0 \\ 0 \end{array} \right]
\end{displaymath}

The other components of Eqn. \ref{EOM} are: 
\begin{displaymath}
\vec{\omega}_{_{N/O_N}} \times (\vec{\omega}_{_{N/O_N}} \times \vec{r}_{_{TMD_X/P_N}}) = x_{_{TMD_X/P_N}} \left[ \begin{array}{c} -(\dot{\phi}_{_{N/O_N}}^2 + \dot{\psi}_{_{N/O_N}}^2) \\ \dot{\theta}_{_{N/O_N}}\dot{\phi}_{_{N/O_N}} \\  \dot{\theta}_{_{N/O_N}}\dot{\psi}_{_{N/O_N}} \end{array} \right]
\end{displaymath}
\begin{displaymath}
2\vec{\omega}_{_{N/O_N}} \times \dot{\vec{r}}_{_{TMD_X/P_N}} = \dot{x}_{_{TMD_X/P_N}} \left[ \begin{array}{c} 0 \\ 2\dot{\psi}_{_{N/O_N}} \\ -2\dot{\phi}_{_{N/O_N}} \end{array} \right]
\end{displaymath}
\begin{displaymath}
\vec{\alpha}_{_{N/O_N}} \times \vec{r}_{_{TMD_X/P_N}} = x_{_{TMD_X/P_N}} \left[ \begin{array}{c} 0 \\ \ddot{\psi}_{_{N/O_N}} \\ -\ddot{\phi}_{_{N/O_N}}\end{array} \right]
\end{displaymath}
Therefore $\ddot{x}_{_{TMD_X/P_N}}$ is governed by the equations
\begin{eqnarray}
\ddot{x}_{_{TMD_X/P_N}} =& (\dot{\phi}_{_{N/O_N}}^2 + \dot{\psi}_{_{N/O_N}}^2-\frac{k_x}{m_x}) x_{_{TMD_X/P_N}} - (\frac{c_x}{m_x}) \dot{x}_{_{TMD_X/P_N}} -\ddot{x}_{_{P/O_N}}+a_{_{G_X/O_N}} \label{EOM_Xx}\\ 
&+ \frac{1}{m_x} ( F_{ext_X} + F_{StopFrc_{X}}) \nonumber
\end{eqnarray} 

The forces $F_{Y_{_{TMD_X/O_N}}}$ and $F_{Z_{_{TMD_X/O_N}}}$ are solved noting $\ddot{y}_{_{TMD_X/P_N}} = \ddot{z}_{_{TMD_X/P_N}} = 0$: 
\begin{eqnarray} \label{FyFz_tmdx}
F_{Y_{_{TMD_X/O_N}}} =&  m_x \left(- a_{_{G_Y/O_N}} +\ddot{y}_{_{P/O_N}} + (\ddot{\psi}_{_{N/O_N}} + \dot{\theta}_{_{N/O_N}}\dot{\phi}_{_{N/O_N}}   ) x_{_{TMD_X/P_N}} + 2\dot{\psi}_{_{N/O_N}} \dot{x}_{_{TMD_X/P_N}} \right)  \label{EOM_Xy}\\ 
F_{Z_{_{TMD_X/O_N}}}  =& m_x \left( - a_{_{G_Z/O_N}} +\ddot{z}_{_{P/O_N}} -(\ddot{\phi}_{_{N/O_N}} - \dot{\theta}_{_{N/O_N}}\dot{\psi}_{_{N/O_N}}   ) x_{_{TMD_X/P_N}} - 2\dot{\phi}_{_{N/O_N}} \dot{x}_{_{TMD_X/P_N}} \right)  \label{EOM_Xz}
\end{eqnarray}

\paragraph{TMD\_Y:} 

The external forces $\vec{F}_{_{TMD_Y/P_N}}$ on $TMD_Y$ are given by
 \begin{displaymath} \vec{F}_{_{TMD_Y/P_N}} =  \left[\begin{array}{c}F_{X_{_{TMD_Y/O_N}}} + m_y a_{_{G_X/O_N}}\\ -c_y \dot{y}_{_{TMD_Y/P_N}} - k_y y_{_{TMD_Y/P_N}} + m_y a_{_{G_Y/O_N}} + F_{ext_y} + F_{StopFrc_{Y}} \\ F_{Z_{_{TMD_Y/O_N}}}+ m_y a_{_{G_Z/O_N}}\end{array} \right] \label{TMDY_force} \end{displaymath}
  
$TMD_Y$ is fixed to frame $N$ in the $x$ and $z$ directions so that
\begin{displaymath}
{r}_{_{TMDYX/P_N}} = \left[ \begin{array}{c} 0 \\ y_{_{TMD_Y/P_N}} \\ 0 \end{array} \right]
\end{displaymath}
The other components of Eqn. \ref{EOM} are: 
\begin{displaymath}
\vec{\omega}_{_{N/O_N}} \times (\vec{\omega}_{_{N/O_N}} \times \vec{r}_{_{TMD_Y/P_N}}) = y_{_{TMD_Y/P_N}} \left[ \begin{array}{c}  \dot{\theta}_{_{N/O_N}}\dot{\phi}_{_{N/O_N}} \\ -(\dot{\theta}_{_{N/O_N}}^2 + \dot{\psi}_{_{N/O_N}}^2)  \\  \dot{\phi}_{_{N/O_N}}\dot{\psi}_{_{N/O_N}} \end{array} \right]
\end{displaymath}
\begin{displaymath}
2\vec{\omega}_{_{N/O_N}} \times \dot{\vec{r}}_{_{TMD_Y/P_N}} = \dot{y}_{_{TMD_Y/P_N}} \left[ \begin{array}{c} -2\dot{\psi}_{_{N/O_N}} \\ 0 \\ 2\dot{\theta}_{_{N/O_N}} \end{array} \right]
\end{displaymath}
\begin{displaymath}
\vec{\alpha}_{_{N/O_N}} \times \vec{r}_{_{TMD_Y/P_N}} = y_{_{TMD_Y/P_N}} \left[ \begin{array}{c} -\ddot{\psi}_{_{N/O_N}} \\ 0 \\ \ddot{\theta}_{_{N/O_N}}\end{array} \right]
\end{displaymath}

Therefore $\ddot{y}_{_{TMD_Y/P_N}}$ is governed by the equations
\begin{eqnarray} 
\ddot{y}_{_{TMD_Y/P_N}} =& (\dot{\theta}_{_{N/O_N}}^2 + \dot{\psi}_{_{N/O_N}}^2-\frac{k_y}{m_y}) y_{_{TMD_Y/P_N}} - (\frac{c_y}{m_y}) \dot{y}_{_{TMD_Y/P_N}} -\ddot{y}_{_{P/O_N}} + a_{_{G_Y/O_N}} \label{EOM_Yy}\\ 
&+ \frac{1}{m_y} (F_{ext_Y} + F_{StopFrc_{Y}}) \nonumber 
\end{eqnarray}
The forces $F_{X_{_{TMD_Y/O_N}}}$ and $F_{Z_{_{TMD_Y/O_N}}}$ are solved noting $\ddot{x}_{_{TMD_Y/P_N}} = \ddot{z}_{_{TMD_Y/P_N}} = 0$:
\begin{eqnarray} \label{Fxz_tmdy}
F_{X_{_{TMD_Y/O_N}}} = m_y \left( - a_{_{G_X/O_N}} + \ddot{x}_{_{P/O_N}} - (\ddot{\psi}_{_{N/O_N}} - \dot{\theta}_{_{N/O_N}}\dot{\phi}_{_{N/O_N}}   ) y_{_{TMD_Y/P_N}} - 2\dot{\psi}_{_{N/O_N}} \dot{y}_{_{TMD_Y/P_N}} \right) \label{EOM_Yx}\\
F_{Z_{_{TMD_Y/O_N}}} = m_y \left( - a_{_{G_Z/O_N}} + \ddot{z}_{_{P/O_N}} + (\ddot{\theta}_{_{N/O_N}} + \dot{\phi}_{_{N/O_N}}\dot{\psi}_{_{N/O_N}}   ) y_{_{TMD_Y/P_N}} + 2\dot{\theta}_{_{N/O_N}} \dot{y}_{_{TMD_Y/P_N}} \label{EOM_Yz}\right) 
\end{eqnarray}

\subsection{State Equations} 
\paragraph{Inputs:}
The inputs are the nacelle linear acceleration and angular position, velocity and acceleration:
\begin{displaymath}
\vec{u} = \left[ \begin{array}{c} \ddot{\vec{r}}_{_{P/O_G}} \\ \vec{R}_{_{N/G}} \\ \vec{\omega}_{_{N/O_G}} \\ \vec{\alpha}_{_{P/O_G}}\end{array} \right]
\Rightarrow \left[ \begin{array}{c} \ddot{\vec{r}}_{_{P/O_N}} \\ \vec{\omega}_{_{N/O_N}} \\ \vec{\alpha}_{_{N/O_N}}\end{array} \right]
 = \left[ \begin{array}{c} \vec{R}_{_{N/G}} \ddot{\vec{r}}_{_{P/O_G}} \\ \vec{R}_{_{N/G}} \vec{\omega}_{_{N/O_G}} \\ \vec{R}_{_{N/G}} \vec{\alpha}_{_{P/O_G}}\end{array} \right]
\end{displaymath} 
\paragraph{States:}
The states are the position and velocity of the TMDs along their respective DOFS in the Nacelle reference frame:

\begin{displaymath}
\vec{R}_{_{TMD/P_N}} = \left[ \begin{array}{c} x \\ \dot{x} \\ y\\ \dot{y} \end{array} \right]_{_{TMD/P_N}} 
= \left[ \begin{array}{c} {x}_{_{TMD_X/P_N}} \\ \dot{x}_{_{TMD_X/P_N}} \\ {y}_{_{TMD_Y/P_N}}\\ \dot{y}_{_{TMD_Y/P_N}} \end{array} \right]
\end{displaymath}

The equations of motion can be re-written as a system of non-linear first-order equations of the form

\begin{displaymath}
\dot{\vec{R}}_{_{TMD}} = A \vec{R}_{_{TMD}} + B 
\end{displaymath} where
\begin{displaymath}
A(\vec{u}) = \left[ \begin{array}{cccc}
0&1&0&0 \\
(\dot{\phi}_{_{P/O_N}}^2 + \dot{\psi}_{_{P/O_N}}^2-\frac{k_x}{m_x}) & - (\frac{c_x}{m_x})&0& 0 \\
0&0&0&1 \\
0& 0 & (\dot{\theta}_{_{P/O_N}}^2 + \dot{\psi}_{_{P/O_N}}^2-\frac{k_y}{m_y}) & - (\frac{c_y}{m_y})  \\ 
\end{array} \right]
\end{displaymath} and

\begin{displaymath}
B(\vec{u}) = \left[ \begin{array}{c}0 \\ -\ddot{x}_{_{P/O_N}}+a_{_{G_X/O_N}} + \frac{1}{m_x} ( F_{ext_X} + F_{StopFrc_{X}}) \\ 0 \\ -\ddot{y}_{_{P/O_N}}+a_{_{G_Y/O_N}} + \frac{1}{m_y} (F_{ext_Y}+ F_{StopFrc_{Y}})  \end{array} \right]
\end{displaymath}

The inputs are coupled to the state variables, resulting in A and B as $f(\vec{u})$. 
\subsection{Outputs}
The output vector $\vec{Y}$ is
\begin{displaymath}
\vec{Y} = \left[ \begin{array}{c} \vec{F}_{_{P_G}} \\ \vec{M}_{_{P_G}} \end{array} \right]
\end{displaymath}
The output includes reaction forces corresponding to $F_{Y_{_{TMD_X/O_N}}}$, $F_{Z_{_{TMD_X/O_N}}}$, $F_{X_{_{TMD_Y/O_N}}}$, and $F_{Z_{_{TMD_Y/O_N}}}$ from Eqns. \ref{EOM_Xy}, \ref{EOM_Xz}, \ref{EOM_Yx}, and \ref{EOM_Yz}. The resulting forces $\vec{F}_{_{P_G}}$ and moments $\vec{M}_{_{P_G}}$ acting on the nacelle are 
\begin{align*}
\vec{F}_{_{P_G}} = R^T_{_{N/G}} & \left[ \begin{array}{c} k_x {x}_{_{TMD/P_N}} + c_x \dot{x}_{_{TMD/P_N}} - F_{StopFrc_{X}} - F_{ext_x} - F_{X_{_{TMD_Y/O_N}}}\\ 
k_y {y}_{_{TMD/P_N}} + c_y \dot{y}_{_{TMD/P_N}} - F_{StopFrc_{Y}} - F_{ext_y} - F_{Y_{_{TMD_X/O_N}}} \\ 
-F_{Z_{_{TMD_X/O_N}}}-F_{Z_{_{TMD_Y/O_N}}} \end{array} \right] 
\end{align*}
and 
\begin{displaymath}
\vec{M}_{_{P_G}} = R^T_{_{N/G}} \left[ \begin{array}{c} M_{_X} \\ M_{_Y} \\ M_{_Z} \end{array} \right]_{_{N/N}} = R^T_{_{N/G}} \left[ \begin{array}{c} -(F_{Z_{_{TMD_Y/O_N}}} ) y_{_{TMD/P_N}} \\
(F_{Z_{_{TMD_X/O_N}}})  x_{_{TMD/P_N}} \\ (-F_{Y_{_{TMD_X/O_N}}})x_{_{TMD/P_N}} + ( F_{X_{_{TMD_Y/O_N}}}) y_{_{TMD/P_N}}  \end{array} \right]
\end{displaymath}

\paragraph{Stop Forces}
The extra forces $F_{StopFrc_{X}}$ and $F_{StopFrc_{Y}}$ are added to output forces in the case that the movement of TMD\_X or TMD\_Y exceeds the maximum track length for the mass. Otherwise, they equal zero. The track length has limits on the upwind (UW) and downwind (DW) ends in the $x$ direction (X\_UWSP and X\_DWSP), and the positive and negative lateral ends in the $y$ direction (Y\_PLSP and Y\_NLSP). If we define a general maximum and minimum displacements as $x_{max}$ and $x_{min}$, respectively, the stop forces have the form 
\begin{displaymath}
F_{StopFrc} = -\left\{
     \begin{array}{lr}
      k_S \Delta x  & : ( x > x_{max} \wedge \dot{x}<=0) \vee ( x < x_{min} \wedge \dot{x}>=0)\\
       k_S \Delta x + c_S \dot{x} & : ( x > x_{max} \wedge \dot{x}>0) \vee ( x < x_{min} \wedge \dot{x}<0)\\
       0 & : $otherwise$
     \end{array}
   \right.
\end{displaymath}
where $\Delta x$ is the distance the mass has traveled beyond the stop position and $k_S$ and $c_S$ are large stiffness and damping constants. 

%
%
%
%
%
%
%
%
%

\section{Code Modifications}
The TMD function is submodule called in ServoDyn. In addition to references in ServoDyn.f90 and ServoDyn.txt, new files that contain the TMD module are listed below. 
\subsection{New Files}
\begin{itemize}
\item TMD.f90 : TMD module
\item TMD.txt : registry file 
\subitem include files, inputs, states, parameters, and outputs shown in Tables \ref{tbl2} and \ref{tbl1} 
\item TMD\_Types.f90 : automatically generated

\end{itemize}
\subsection{Variables}
\begin{table}
\centering
\renewcommand{\arraystretch}{1.2}\addtolength{\tabcolsep}{.5pt}
\begin{tabularx}{\textwidth}{llllll}
InitInput & 			Input u & 							Parameter p & 					State x & 					Output y \\ \hline
InputFile & 			$\ddot{\vec{r}}_{_{P/O_G}}	$ & 	$m_x$ & 						$\vec{tmd_x}$ 			& 	Mesh \\
Gravity & 				$\vec{R}_{_{N/O_G}} 		  	$ & $c_x$ & 						 	& 						 \\
$\vec{r}_{_{N/O_G}}$&	$\vec{\omega}_{_{N/O_G}}	$ & 	$k_x$ &								\\
&						$\vec{\alpha}_{_{P/O_G}} 	$ & 	$m_y$ &										\\
 && 														$c_y$ &								\\
& 						&									$k_y$  &\\ 
& & 														$K_S = \left[ k_{SX}\hspace{1em}k_{SY}\right]$ & \\
&& 															$C_S = \left[c_{SX}\hspace{1em}c_{SY}\right]$ & \\
&& 															$P_{SP} = \left[ X_{DWSP} \hspace{1em}Y_{PLSP}\right]$ \\
&& 															$P_{SP} = \left[ X_{UWSP} \hspace{1em}Y_{NLSP}\right]$ & \\
&& 															$F{ext}$ & \\
&& 															$Gravity$ & \\
&&TMDX\_DOF & \\
&&TMDY\_DOF & \\ 
&&$X_{DSP}$ \\
&&$Y_{DSP}$ \\ \hline
\end{tabularx}
\caption{Summary of field definitions in the TMD registry. Note that state vector $\vec{tmd_x}$ corresponds to $\vec{R}_{_{TMD/P_N}}$, and that the outputs $\vec{F}_{_{P_G}}$ and $\vec{M}_{_{P_G}}$ are contained in the MeshType object (y.Mesh). $X_{DSP}$ and $Y_{DSP}$ are initial displacements of the TMDs.}
\label{tbl2}
\end{table}
The input, parameter, state and output definitions are summarized in Table \ref{tbl2}. The inputs from file are listed in Table \ref{tbl1}.
\begin{table}
\small
\centering
\renewcommand{\arraystretch}{1.2}\addtolength{\tabcolsep}{.5pt}
\begin{tabularx}{\textwidth}{p{3cm} p{2cm} X}
Field Name & Field Type & Description \\ \hline
TMD\_CMODE & int  & Control Mode (1:passive, 2:active) \\
TMD\_X\_DOF & logical  & DOF on or off\\
TMD\_Y\_DOF & logical  & DOF on or off\\
TMD\_X\_DSP & real  & TMD\_X initial displacement\\
TMD\_Y\_DSP & real  & TMD\_Y initial displacement\\
TMD\_X\_M & real  & TMD mass \\
TMD\_X\_K & real  & TMD stiffness \\
TMD\_X\_C &  real  & TMD damping \\
TMD\_Y\_M & real  & TMD mass \\
TMD\_Y\_K & real  & TMD stiffness \\
TMD\_Y\_C & real  & TMD damping \\
TMD\_X\_DWSP & real  & DW stop position (maximum X mass displacement) \\
TMD\_X\_UWSP & real  & UW stop position (minimum X mass displacement) \\
TMD\_X\_K\_SX & real & stop spring stiffness \\
TMD\_X\_C\_SX & real & stop spring damping \\
TMD\_Y\_PLSP & real  & positive lateral stop position (maximum Y mass displacement) \\
TMD\_Y\_NLSP & real  & negative lateral stop position (minimum Y mass displacement) \\
TMD\_Y\_K\_S & real & stop spring stiffness \\
TMD\_Y\_C\_S & real & stop spring damping \\
TMD\_P\_X & real & x origin of P in nacelle coordinate system \\
TMD\_P\_Y & real & y origin of P in nacelle coordinate system \\
TMD\_P\_Z & real & z origin of P in nacelle coordinate system \\ \hline
\end{tabularx}
\caption{Data read in from TMDInputFile.}
\label{tbl1}
\end{table}
\section{Acknowledgements}
The authors would like to thank Dr. Jason Jonkman for reviewing this manual. 

\bibliographystyle{plain}
\interlinepenalty=900000000 
\bibliography{TMD}
\end{document}